\documentclass[conference]{IEEEtran}
\IEEEoverridecommandlockouts
\usepackage{cite}
\usepackage{amsmath,amssymb,amsfonts}
\usepackage{algorithmic}
\usepackage{graphicx}
\usepackage{textcomp}
\usepackage{xcolor}
\def\BibTeX{{\rm B\kern-.05em{\sc i\kern-.025em b}\kern-.08em
    T\kern-.1667em\lower.7ex\hbox{E}\kern-.125emX}}

\usepackage{multirow}
\usepackage{url} 
\usepackage{hyperref}
\usepackage{booktabs} 

\begin{document}

\title{Synthetic Data Generation with Large Language Models for Personalized Community Question Answering\\
}


\author{\IEEEauthorblockN{Marco Braga}
\IEEEauthorblockA{\textit{Department of Informatics, Systems and Communication}\\
\textit{University of Milano-Bicocca}\\
Milan, Italy\\
m.braga@campus.unimib.it}
\and
\IEEEauthorblockN{Pranav Kasela}
\IEEEauthorblockA{\textit{Department of Informatics, Systems and Communication}\\
\textit{University of Milano-Bicocca}\\
Milan, Italy\\
p.kasela@campus.unimib.it}
\and
\IEEEauthorblockN{Alessandro Raganato}
\IEEEauthorblockA{\textit{Department of Informatics, Systems and Communication}\\
\textit{University of Milano-Bicocca}\\
Milan, Italy\\
alessandro.raganato@unimib.it}
\and
\IEEEauthorblockN{Gabriella Pasi}
\IEEEauthorblockA{\textit{Department of Informatics, Systems and Communication}\\
\textit{University of Milano-Bicocca}\\
Milan, Italy\\
gabriella.pasi@unimib.it}
}

\maketitle

\begin{abstract}
Personalization in Information Retrieval (IR) is a topic studied by the research community since a long time. However, there is still a lack of datasets to conduct large-scale evaluations of personalized IR; this is mainly due to the fact that 
collecting and curating high-quality
user-related information
requires significant costs and time investment.
Furthermore, the creation of datasets for Personalized IR (PIR) tasks is 
affected by both privacy concerns and the need for 
accurate
user-related data, which are often not publicly available. Recently, researchers have started to explore the use of Large Language Models (LLMs) to generate synthetic datasets, which is a possible solution to generate data for low-resource tasks. In this paper, we investigate the potential of Large Language Models (LLMs) for generating synthetic documents to train an IR system for a 
Personalized Community Question Answering task. To study the effectiveness of IR models fine-tuned on LLM-generated data,
we introduce a new dataset, named Sy-SE-PQA.
We build Sy-SE-PQA based on an existing dataset, SE-PQA \footnote{\url{https://zenodo.org/records/10679181}}, which consists of questions and answers posted on the popular StackExchange communities. Starting from questions in SE-PQA, we generate synthetic answers using different prompt techniques and LLMs.
Our findings suggest that LLMs have high potential in generating data tailored to users' needs. The synthetic data can replace human-written training data, even if the generated data may contain incorrect information.
The code is publicly available\footnote{\url{https://github.com/pkasela/SY_SE-PQA}}.
\end{abstract}


\section{Introduction} 
The topic of Personalized Information Retrieval (PIR) has been studied by researchers since a long time \cite{speretta2005personalized, borisov2016context, wu2017ensemble, Bassani_cikm, bassani-etal-2024-denoising, braga2023personalization, braga_sigir}. The goal of PIR is to produce an output tailored to a specific user (or group of users) by leveraging the user's interests and preferences 
, which are captured by
gathering user-related information. 
The performance of PIR models based on neural models significantly depends on the quality of the training data; this is 
actually still a challenge, due to the scarcity of large-scale, publicly available datasets that include detailed user information, which hinders the training of effective personalized neural models. 
Existing datasets like the AOL query log \cite{pass2006picture}, the Yandex query log\footnote{\href{https://www.kaggle.com/c/yandex-personalized-web-search-challenge}{Yandex Query Log}} and the CIKM Cup 2016 dataset,\footnote{\href{https://competitions.codalab.org/competitions/11161}{CIKM Cup 2016}} even if commonly used, have privacy concerns or limitations due to anonymization \cite{barbaro2006face}.
Recently, a new dataset, called SE-PQA \cite{kasela2024se}, has contributed to fill this gap; it is built by collecting questions and answers from StackExchange, the well-known community Question Answering (cQA) platform. The goal of SE-PQA is to ease the design of personalized Question Answering (QA) systems adapted to an IR task, where the question is seen as a query, and the answers are retrieved from a pool of answers. SE-PQA contains questions (i.e. queries) from 50 different communities, which can be categorized under the large umbrella of humanistic communities. 

\begin{figure}[ht]
    \centering
    \includegraphics[width=.9\linewidth]{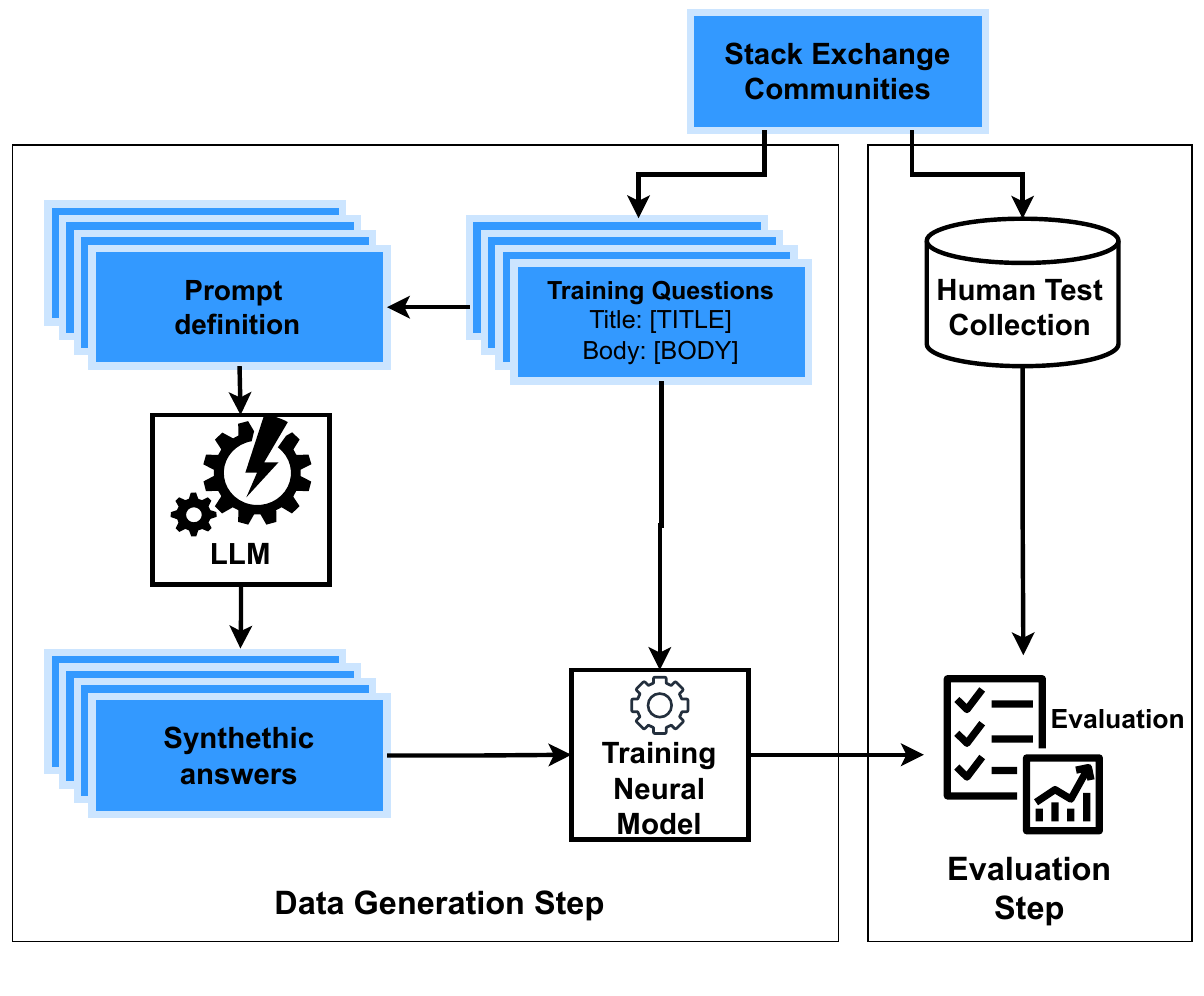}
    \caption{Illustration of the process of Data Generation and Evaluation. In the Data Generation step, we generate the synthetic answers to questions sampled from StackExchange. Then we train neural retrieval models where the queries are the questions and the documents the synthetic answers. Finally, in the Evaluation step, we evaluate the retrieval models on a human-written test set.}
    \label{fig:pipeline}
\end{figure}

In the last few years, as a possible solution to automatically create data for low-resource tasks, researchers have explored the use of LLMs to generate synthetic datasets.
LLMs, such as GPT \cite{brown2020language}, have shown capabilities in generating synthetic data for different tasks such as Text Classification \cite{salemi2023lamp}, Relation Extraction in the medical domain \cite{tang2023does}, Dialogue Systems \cite{Perttudialogue23, aliannejadi2024trec} and Information Retrieval (IR) \cite{bonifacio2022inpars, askari_gpt}. 
However, little effort has been devoted to generating documents tailored to users' needs
for an IR task.
In this work, we investigate the application of Large Language Models (LLMs) to generate synthetic data for personalized community question answering (cQA). As shown in Figure \ref{fig:pipeline}, the process begins with the collection of user-related information and questions from the SE-PQA dataset. We then employ recent LLMs, namely, GPT-3.5 \cite{brown2020language} and Phi-3 \cite{abdin2024phi}, to generate synthetic answers, tailored to individual user interests.
To evaluate the effectiveness of the LLM-generated data, we train neural retrieval models using
these data.
The performance of these models is evaluated on a human-annotated test set, allowing us to assess how effectively the synthetic data enhance personalized information retrieval compared to models trained on traditional, human-authored data. 
A known challenge with LLMs is their tendency to generate inaccurate or misleading information, a phenomenon often referred to as hallucination. To address this issue, we conducted a manual evaluation of the synthetic answers to identify and assess the extent of hallucinations present in the proposed synthetic dataset.
We observe that training neural retrieval models using LLM-generated data achieves better performance compared to the human-annotated data, which shows the potential of using LLMs to generate training data for a personalized question-answering task. Furthermore, we find that more than $35\%$ of generated answers contain hallucinations, showing that training neural IR models with factually true answers does not seem mandatory.  

To summarize, our main contributions in this work are three-fold:
\begin{enumerate}
    \item We release the Sy-SE-PQA dataset, which is a synthetic dataset for a personalized question-answering task. This dataset contains answers to 100k questions from StackExchange. 
    \item We present a comprehensive evaluation, including fine-tuning DistillBERT \cite{sanh2019distilbert} on both human-written and LLMs-generated answers, evaluating then their performance on a human-written test set. 
    \item We analyze the data diversity concerning different LLMs and prompts. Furthermore, we manually analyse the generated answers to verify that the generated training data may not be factually accurate.  
\end{enumerate}

The paper is organised as follows. Section \ref{RW} discusses the relevant Related Works. Section \ref{methodology} introduces in detail the procedure employed to generate synthetic data for personalized cQA task, while Section \ref{Results} discusses the results of our experimental evaluation. Finally, Section \ref{subsec:exploratory_analysis} discusses the data diversity concerning different LLMs and prompts.      


\section{Related Works}
\label{RW}
In this section, we describe the related works for both Large Language Models (LLMs) and synthetic data generation.

\subsection{Large Language Models}

Based on the Transformer architecture \cite{vaswani2017attention}, Large Language Models (LLMs) have shown remarkable performances in the Natural Language Processing (NLP) field. This progress was further advanced by OpenAI's GPT series, including GPT-2 \cite{radford2019language} and GPT-3.5 (ChatGPT) \cite{brown2020language}. These models are 
characterized by their ability to generate high-quality, human-like text \cite{clark2021all,dou2022gpt,zhou2023synthetic}, and 
exhibit proficiency in fundamental reasoning \cite{wei2021finetuned}, translation \cite{brown2020language}, scientific synthetic data generation \cite{hamalainen2023evaluating} and code generation \cite{mcnutt2023design}. Recently, Microsoft introduced a family of small language models, called Phi-3 \cite{abdin2024phi}. The version Phi-3-mini, which has 3.8 billion parameters, has an overall comparable performance as GPT-3.5 \cite{abdin2024phi}, despite its reduced size. For instance, Phi-3-mini achieves competitive performance across a wide range of tasks, such as mathematical problem-solving dataset \cite{hendrycks2021measuring} and multi-turn questions answering \cite{zheng2024judging}, compared with closed LLMs such as GPT-3.5.


\subsection{Synthetic Data Generation}

Recent advancements in Generative Artificial Intelligence have motivated researchers to explore the potential of generative models to create synthetic data for training machine learning models, especially in Computer Vision and NLP tasks. In the context of this paper, we focus on the generation of textual data. Researchers have explored the ability of LLMs to generate data for various text classification tasks, with mixed results regarding their effectiveness \cite{kumar2020data,yoo2021gpt3mix,aggarwal2022entity,chen2022weakly,gao2022self,meng2022generating,sahu2022data,ye2022zerogen,chung2023increasing}. 
Puri et al. \cite{puri-etal-2020-training} explore question and answer generation as a data augmentation method to enhance QA models, addressing the challenge of limited human-labelled data. 
The authors point out that the synthetic data alone overcome the performance of the original training set in the Exact Match (EM) and F1 scores. 
Li et al. \cite{li-etal-2023-synthetic} investigate the effectiveness of synthetic data generated by LLMs in training text classification models across tasks with varying levels of subjectivity. They compare models trained on real-world data with those trained on LLM-generated data under zero-shot and few-shot settings. Their findings reveal that while models trained on synthetic data perform comparably on low-subjectivity tasks, e.g. news and topic classification, there is a notable performance drop on high-subjectivity tasks, such as sarcasm detection. 
Regarding the generation of synthetic data for IR tasks, LLMs can be used to generate synthetic queries \cite{bonifacio2022inpars} or documents \cite{askari_gpt}. In
\cite{wang2023query2doc} the authors propose to generate
documents by few-shot prompting LLMs and concatenate them with the original query to form a new query.  

We focus our work on personalization tasks, which
aim to tailor a model's outcomes for a specific user (or group of users) based on the knowledge of her/his interests and behaviour. For example, as the use of LLMs in real-world applications evolves, recent work has highlighted the impact and concerns associated with personalizing LLMs \cite{Huang_user_nlp}. Salemi et al. \cite{salemi2023lamp} introduce the LaMP benchmark, a synthetic dataset designed to evaluate the ability of LLMs to produce personalized outcome. LaMP encompasses seven distinct tasks, divided into three Text Classification tasks (Citation Identification, Movie Tagging and Product Rating) and four Text Generation tasks (News Headline, Scholarly Title, Email Subject and Tweet Paraphrasing). Chan et al. \cite{chan2024scalingsyntheticdatacreation} use text data from the Web as input to an LLM to generate a textual description of a user who likely wrote the input text. Then, the authors use the user description as input to an LLMs, such as GPT-4 \cite{achiam2023gpt}, to create synthetic data, including Maths and Logical reasoning problems, tailored to the specific user.

We focus our work on a specific personalized task, \textit{Personalized Information Retrieval} (PIR). Personalization in IR \cite{liu2020personalization} aims to tailor search results to specific users to overcome the one-size-fits-all behaviour of search engines. Two main techniques in the PIR field are query expansion and results re-ranking. Personalized query expansion \cite{bassani2023personalized} relies on user-related documents to extract expansion terms from the user's vocabulary. The original user's query is augmented with the new expansion terms to improve the system's effectiveness. In results re-ranking, a retrieval system’s component, called the first-stage retriever, retrieves a ranked list of documents in response to a search query. Then, another component, the re-ranker, computes new relevance scores for the initially retrieved documents leveraging additional information. Given the set of documents provided by the first-stage retriever, a Personalized Re-Ranker \cite{kasela2024se} orders the documents by leveraging user-related information, usually comparing the user’s interests and preferences with the topics covered by the
documents. 
Working with personalized data and information presents significant challenges related to privacy and anonymization \cite{barbaro2006face}, which makes the possibility of generating synthetic data based on user-related information useful.
Aliannejadi et al. \cite{aliannejadi2024trec} propose a test collection, the so-called TREC Interactive Knowledge Assistance Track (iKAT), designed to evaluate Conversational Search Agents through a comprehensive dataset of personalized dialogues. The collection features $36$ dialogues spanning $20$ topics, each integrated with a personal text knowledge base that defines users' personas. The iKAT aims to address challenges in conversational systems, including context-dependency, personalization, and mixed-initiative dialogues. The collection utilizes the ClueWeb22-B dataset \cite{overwijk2022clueweb22} for passage retrieval, with human-annotated dialogues to ensure high-quality evaluations. 

\section{Methodology}
\label{methodology}
In this section, we begin by outlining the process of generating synthetic data for our PIR task 
(Section \ref{datagen}). 
Following the dataset generation, we proceed to detail the models employed to evaluate the effectiveness of our generated data (Section \ref{Baselines}).

\subsection{Dataset Generation}
\label{datagen}

\begin{figure*}[ht]
    \centering
    \includegraphics[width=\linewidth]{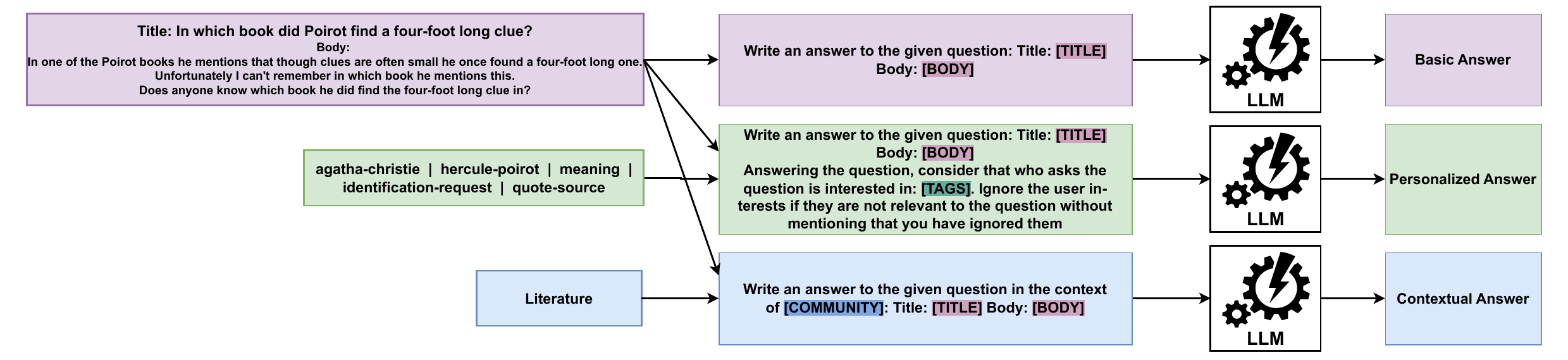}
    \caption{Example of a question with the three prompts utilized.}
    \label{fig:prompt}
\end{figure*}

Our Sy-SE-PQA dataset is based on the personalized version of the SE-PQA dataset \cite{kasela2024se}, which contains over 200k training questions sampled from 50 different StackExchange communities. 
Each of these communities represents a different domain of knowledge or interest, ranging from gardening to linguistics, in which a query, consists of a Title and a Body.
The title summarizes the question, typically highlighting the core issue or query in a brief, direct manner. It acts as a quick reference point for users browsing through questions. 
The Body, on the other hand, provides the necessary context, elaborating on the details surrounding the query.
This may include background information, specific examples, or clarifications needed to fully understand the question. Together, the Title and the Body form a comprehensive representation of the user’s information need.
Furthermore, when posting a question, users have the option to assign tags to them. The tags are selected from a predefined set and serve multiple purposes: they categorize the question, making it easier to find other users interested in similar topics, and they also provide an additional layer of context that can be leveraged by models to generate more targeted and relevant responses. On average, each question has 2 to 3 tags associated with them. 
To simulate different scenarios and user preferences, we generate three different type of answers for each question. These variations of answers differ in terms of the level of personalization and contextual information provided to the large language models (LLMs). 
These variations allow us to assess the adaptability and relevance of the generated answers across different contexts:
    \paragraph{Basic Answer Generation} The first approach is the generation of basic, straightforward answers. In this scenario, the model is tasked with producing an answer based solely on the content provided in the question's Title and Body, without any additional context or personalization. The prompt used to generate these synthetic answers is as follows:
    
    \textit{``Write an answer to the given question: Title: [TITLE] Body: [BODY].''}
    
    \paragraph{Personalized Answer Generation} The second approach introduces an element of personalization into the answer generation process. Since a score based on the tags associated with each user has been successfully used to rank search results in a personalized way \cite{kasela2024se}, we achieve personalization 
    by incorporating data about the user’s interests inferred from the five tags they have most frequently used in their previous interactions. 
    As described in \cite{kasela2024se}, almost all questions contain up to five tags each, and we followed this insight using the same number of tags as user-related interests.
    The modified prompt for generating personalized answers is:
    
    \textit{``Write an answer to the given question: Title: [TITLE] Body: [BODY].\\
    Answering the question, consider that who asks the question is interested in: [TAGS]. Ignore the user interests if they are not relevant to the question without mentioning that you have ignored them.''}

    By including this personalization layer, the model is guided to consider the user's potential interests or preferences while formulating its response. This is particularly useful in scenarios where a user's history of engagement with certain topics can influence the type of response they are seeking. 
    The instruction to ``ignore the user interests if they are not relevant'' ensures that the model remains focused on providing accurate and pertinent answers, without being unduly swayed by irrelevant personal data.
    
    \paragraph{Contextual Answer Generation} The third approach focuses on incorporating contextual information about the community in which the question was posted. StackExchange communities are specialized forums where users with similar interests or expertise gather to discuss and resolve issues within a specific domain. By embedding this community context into the prompt, the model can generate answers that are more aligned with  the specific community. The prompt is as follows:
    
    \textit{``Write an answer to the given question in the context of [COMMUNITY]: Title: [TITLE] Body: [BODY].''}

    Including the community context allows the model to tap into domain-specific knowledge and adjust its responses accordingly. For instance, a question about historical figures posted in a History community might be answered differently than the same question posted in a general knowledge or movies community. 
    This prompt, actually, does not aim at generating an answer personalized according to the individual user, rather it aims at adapting the generated response to a specific group of users, which are the users who wrote in the community.
For example, a user might inquire about the identity of a historical figure represented in a specific movie scene. Instead of posting this question in a Movies community, where the emphasis would be on cinematic details or trivia, the user may opt to ask in the History community. This decision would suggest that the user is more interested in the historical significance of the figure rather than its role in the movie.
By understanding and utilizing the community context, the model can generate an answer that delves into the historical aspects, thereby better aligning with the user’s underlying intent and providing a more meaningful and satisfactory response.

To carry out the generation of these answers we rely on one closed source model: 
\textit{GPT-3.5 Turbo} \cite{brown2020language}, and two open-source models: \textit{Phi-3-4k mini}\footnote{\href{https://huggingface.co/microsoft/Phi-3-mini-4k-instruct}{microsoft/Phi-3-mini-4k-instruct}} (3.82 B) \cite{abdin2024phi}, and \textit{Phi-3-4k-medium}\footnote{\href{https://huggingface.co/microsoft/Phi-3-medium-4k-instruct}{microsoft/Phi-3-medium-4k-instruct}} (14 B) \cite{abdin2024phi}.
We selected the two Phi models as the open-source LLMs due to their state-of-the-art performance across a variety of natural language processing tasks. At the time of this study, these models achieve high performance in areas such as language understanding, math, code, long context and logical reasoning \cite{abdin2024phi}. 

Given the high computational costs associated with generating large volumes of data, we imposed a limit on the number of questions sampled from each community in the SE-PQA dataset. 
Specifically, we restricted this number to at most $3,000$ questions per community. 
If the community has less than $3,000$ questions, we sample all of them. 
This ensures a broad yet manageable variance of questions for our analysis.
Despite this restriction, the generated data contains approximately  100k answers for each prompt with all three LLMs.
For Phi-3 mini and medium, we use a temperature of 1, and a maximum number of tokens generated equal to 500, as suggested by the Phi-3 documentation.\footnote{\url{https://huggingface.co/microsoft/Phi-3-mini-4k-instruct}} To make a fair comparison, we use the same hyperparameter for GPT-3.5.

\begin{table*}[ht]
    \centering
    \caption{Example of hallucinations in the Movies community
    }
    \resizebox{\textwidth}{!}{
\begin{tabular}{llp{.9\textwidth}}
Question Title & \multicolumn{2}{c}{Identify the movie where a cartoonist turns into an animation}           \\ \hline
Question Body & 
 \multicolumn{2}{c}{I remember a movie, probably italian or some similar latin language, where a cartoonist is slowly transformed into an animated character. [...]}        \\ \hline
Model & Prompt type & Answer \\ \hline
Human written &  &  I'm pretty sure you're talking about \textbf{Volere Volare (1991)}, an italian movie where the main character has a date with a girl, but he's so nervous that he transformes into an animated character.      \\ \hline
{\multirow{4}{*}{GPT 3.5}}  
      & \textit{Personalized}          &   The movie you are thinking of is likely \textbf{"Monkeybone"} which was released in 2001. [...] \\
  
  & \textit{Basic}           &   The movie you are referring to is likely \textbf{"Monkeybone"} a 2001 American dark comedy film starring Brendan Fraser. [...] ",
  \\   
      & \textit{Contextual}            &     The movie you are referring to is likely \textbf{"Il Lupo perde il pelo"}, also known as "The Wolf Loses His Fur". [...]. This Italian film was released in 1987 and features a cartoonist who is transformed into an animated character. [...] \\ \hline

      {\multirow{4}{*}{Phi-mini}}  
      & \textit{Personalized}          &  The movie you are referring to is \textbf{"La vita é bella"} (Life Is Beautiful), directed by Roberto Benigni. However, it's important to note that the plot of "La vita é bella" does not involve a cartoonist turning into an animated character. [...]
      \\ 
  & \textit{Basic}          & [...] The movie you're referring to is likely \textbf{"La vita è bella"} (Life Is Beautiful), directed by Roberto Benigni. [...]
  \\   
      & \textit{Contextual}            &  The movie you're referring to is likely \textbf{"La vita é bella"} (Life Is Beautiful), directed by Roberto Benigni. [...]
      \\ \hline
      {\multirow{4}{*}{Phi-medium}}     
      & \textit{Personalized}          &      The movie you're referring to is likely \textbf{"Lucky and Zorba"} (also known as "Lucky and Zorba: The Coolest Cats in the World"), an Italian animated film released in 1998. [...]
      \\ 
  & \textit{Basic}           &  The movie you're referring to is likely \textbf{"Lucky and Zorba"} [...]. It's an Italian animated film released in 1998. [...]
  \\   
      & \textit{Contextual}            &    The movie you're referring to is likely \textbf{"Lucky and Zorba"} [...]. This Italian animated film was released in 2005, a few years after "Who Framed Roger Rabbit" (1988). [...]
      \\ \hline
\end{tabular}
    }
    \label{tab:confronto_history}

    \end{table*}
    
\subsection{Baselines and Evaluation}
\label{Baselines}

To establish the baseline, in this study, we follow the two-stage ranking architecture, which is the same as utilized in the SE-PQA evaluation \cite{kasela2024se}, aimed at trading-off effectiveness and efficiency by applying two increasingly accurate and computationally expensive ranking models. It aims at selecting for each query a set of candidate documents that are eventually re-ranked by the second, precision-oriented ranker. 
These baselines enable us to assess the performance of our synthetic data generation methods and the various models used in the PIR task. 

The \textit{first stage} is based on \textit{BM25} 
implemented with elasticsearch. 
BM25 hyperparameters, \textit{b} and \textit{$k_1$} are set to be 1 and 1.75, respectively, the same 
used by the original SE-PQA work \cite{kasela2024se}.
In the \textit{second stage} we re-score the initial BM25 ranking by incorporating a deep learning-based approach using \textit{DistilBERT}\footnote{\href{https://huggingface.co/distilbert/distilbert-base-uncased}{distilbert/distilbert-base-uncased}} \cite{sanh2019distilbert}. In all the experiments, the second stage re-ranks the top-100 results retrieved by BM25. 
The document relevance score in the second stage is determined by a convex combination of the scores computed by BM25 and DistilBERT. This combined approach allows us to harness the strengths of both traditional information retrieval techniques and modern neural methods. 
To determine the optimal weighting between these two components, we perform a grid search over the interval $[0,1]$, with a step size of 0.1. This grid search is conducted on the original SE-PQA validation set to identify the best-performing weights. 
The optimal weights for each model are reported in Tables \ref{tab:phi-mini}, \ref{tab:phi-medium} and \ref{tab:gpt-35}.



We fine-tune DistilBERT specifically for this task, training it for 20 epochs. We choose a batch size of $128$ and a learning rate of $5 \cdot 10^{-6}$. We employ Triplet Margin Loss as our loss function, with in-batch random negatives, 
and a margin of $\gamma=0.5$. 
All the answers not relevant to a question are used as in-batch random negatives, and during the training procedure for each type of answer the model is allowed to see only the answers generated by the same type of prompts.
We use AdamW as the optimizer and set the random seed to $42$ to ensure reproducibility across all our experiments. 
We train and evaluate our models on a single A100 GPU. 

\section{Results}
\label{Results}
In this section, we present the outcomes of our experiments, evaluated using
P@1, NDCG@3, NDCG@10 and MAP@100 as evaluation metrics. 
These metrics are crucial for assessing the precision and relevance of the retrieved documents. All metrics are computed using the ranx library \cite{bassani2022ranx}.
\begin{table}
    \centering
    \caption{Results on SE-PQA test set using models trained on data from Phi-mini.}
    \label{tab:phi-mini}
    \resizebox{\linewidth}{!}{
    \begin{tabular}{llccccc}
        Training & Model & P@1 & NDCG@3 & NDCG@10 & MAP@100 & $\lambda$\\
        \midrule
        - & BM25 &  0.279 & 0.353 & 0.394 &  0.362 & -\\
        
        \midrule

        \multirow{2}{*}{\textit{Real Answer}}
        & DistilBERT &  0.253 & 0.329 & 0.378 &  0.344 & -\\
        & BM25 + DistilBERT & 0.333* & 0.415* & 0.456* &  0.421* & .3\\

        \midrule
        
        \multirow{2}{*}{\textit{Basic}}
        
        & DistilBERT &  0.264 & 0.340 & 0.388 &  0.352 & -\\
        & BM25 + DistilBERT & 0.336* & 0.419* & 0.460* &  0.425* & .3\\

        \midrule
        
        \multirow{2}{*}{\textit{Personalized}}

        & DistilBERT &  0.299* & 0.374* & 0.418* &  0.384* & -\\
        & BM25 + DistilBERT & 0.345* & 0.426* & 0.465* &  0.431* & .3\\

        \midrule
        
        \multirow{2}{*}{\textit{Contextual}}
        
        & DistilBERT &  0.294* & 0.369* & 0.413* &  0.379* & -\\
        & BM25 + DistilBERT & 0.342* & 0.425* & 0.464* &  0.429* & .3\\

    \end{tabular}
    }
\end{table}
\begin{table}
    \centering
    \caption{Results on SE-PQA test set using models trained on data from Phi-medium.}
    \label{tab:phi-medium}
    \resizebox{\linewidth}{!}{
    \begin{tabular}{llccccc}
        Training & Model & P@1 & NDCG@3 & NDCG@10 & MAP@100 & $\lambda$\\
        \midrule
        - & BM25 &  0.279 & 0.353 & 0.394 &  0.362 & -\\
        
        \midrule

        \multirow{2}{*}{\textit{Real Answer}}
        & DistilBERT &  0.253 & 0.329 & 0.378 &  0.344 & -\\
        & BM25 + DistilBERT & 0.333* & 0.415* & 0.456* &  0.421* & .3\\

        \midrule
        
        \multirow{2}{*}{\textit{Basic}}
        
        & DistilBERT &  0.299* & 0.373* & 0.418* &  0.383* & -\\
        & BM25 + DistilBERT & 0.344* & 0.425* & 0.465* &  0.430* & .3\\
        
        \midrule
        
        \multirow{2}{*}{\textit{Personalized}}
        
        & DistilBERT &  0.299* & 0.374* & 0.419* &  0.384* & -\\
        & BM25 + DistilBERT & 0.347* & 0.429* & 0.468* &  0.433* & .3\\
        
        \midrule
        
        \multirow{2}{*}{\textit{Contextual}}
        
        & DistilBERT &  0.301* & 0.375* & 0.419* &  0.385* & -\\
        & BM25 + DistilBERT & 0.338* & 0.423* & 0.463* &  0.427* & .4\\
        
    \end{tabular}
    }
\end{table}
Tables \ref{tab:phi-mini}, \ref{tab:phi-medium} and \ref{tab:gpt-35} summarize the results for the models trained using data from Phi-mini, Phi-medium and GPT-3.5, respectively.
In these tables, asterisks (*) indicate statistically significant improvements over the BM25 method, determined using a Bonferroni-corrected two-sided paired Student's t-test with 99\% confidence. The $\lambda$ column shows the optimized weight for BM25 during the second stage, with the weight for the neural model being $1-\lambda$. 
We evaluate the models' effectiveness on the validation and test set from SE-PQA 
and thus their ability to retrieve human responses. 

\begin{table}
    \centering
    \caption{Results on SE-PQA test set using models trained on data from GPT-3.5.}
    \label{tab:gpt-35}
    \resizebox{\linewidth}{!}{
    \begin{tabular}{llccccc}
        Training & Model & P@1 & NDCG@3 & NDCG@10 & MAP@100 & $\lambda$\\
        \midrule
        - & BM25 &  0.279 & 0.353 & 0.394 &  0.362 & -\\
        
        \midrule

        \multirow{2}{*}{\textit{Real Answer}}
        & DistilBERT & 0.269 & 0.346 & 0.394 & 0.358 & -\\
        & BM25 + DistilBERT & 0.336* & 0.419* & 0.459* & 0.424* & .3\\

        \midrule
        
        \multirow{2}{*}{\textit{Basic}}

        & DistilBERT & 0.274 & 0.347 & 0.394 & 0.360 & -\\
        & BM25 + DistilBERT & 0.335* & 0.415* & 0.455* & 0.420* & .3\\

        \midrule
        
        \multirow{2}{*}{\textit{Personalized}}
        
        & DistilBERT & 0.263 & 0.333 & 0.380 & 0.347 & -\\
        & BM25 + DistilBERT & 0.325* & 0.406* & 0.448* & 0.413* & .4 \\
        
        \midrule
        
        \multirow{2}{*}{\textit{Contextual}}

        & DistilBERT & 0.275 & 0.346 & 0.392 & 0.359 & -\\
        & BM25 + DistilBERT & 0.331* & 0.411* & 0.452* & 0.417* & .4 \\

    \end{tabular}
    }
\end{table}

To compare performance between models trained on synthetic and real data, we train DistilBERT on human-written data, i.e., the real answers labelled as relevant for the user who wrote the question in the SE-PQA training set. Results of DistilBERT fine-tuned on human-written data are reported in the rows labelled \textit{Real Answers}.
From the Tables \ref{tab:phi-mini}, \ref{tab:phi-medium} and \ref{tab:gpt-35}, we notice that neural re-ranker DistilBERT, trained on synthetic data, outperforms BM25 in a significant way, indicating that the synthetic dataset is suitable for training neural retrieval models. 
Indeed, DistilBERT improvement reaches 24\% in P@1 and 18\% in NDCG@10 over the BM25 model, when fine-tuned on the \textit{Personalized} Answers.
Even without adding contextual or personal information into the prompt, the model trained on \textit{Basic} synthetic data achieves similar performances to the model trained on \textit{Real Answers}.
Furthermore, training the model on \textit{Personalized} answers generated by both Phi mini and medium is able to improve remarkably over the model trained on the real answer. Regarding the data generated with GPT-3.5, it is worth noting that, even if the model trained on \textit{Basic} generated answers achieves the best results compared to the model trained on synthetic data, the performance of all the models are in the same ballpark. Compared with the ones trained on \textit{Real Answer}, models trained on synthetic data showed marginally lower performance. Furthermore, in contrast with the models trained with Phi-generated data, the \textit{Personalized} data generated by GPT-3.5 achieve the lowest performance compared to the other prompts.  




\section{Exploratory Analysis}
\label{subsec:exploratory_analysis}
In this Section, we first evaluate the differences between the answers generated using the three different prompts, and then if the generated answers contained incorrect information. 


\begin{table*}[ht]
    \centering
\begin{minipage}[t]{0.48\linewidth}\centering
\caption{Comparing the diversity of synthetic generated data using different input prompts by applying BLEU score}
\label{Tabella_bleu}
\begin{tabular}{ccc}
\multicolumn{3}{c}{\textbf{Phi-mini}}                                 \\
\multicolumn{1}{c|}{\textit{Prompt type}} & \textit{Contextual} & \textit{Personalized} \\ \hline
\multicolumn{1}{c|}{\textit{Basic}}                & 0.29       & 0.24         \\
\multicolumn{1}{c|}{\textit{Contextual}}           &  -          & 0.23         \\
\multicolumn{3}{c}{\textbf{Phi-medium}}                               \\
\multicolumn{1}{c|}{\textit{Prompt type}} & \textit{Contextual} & \textit{Personalized} \\ \hline
\multicolumn{1}{c|}{\textit{Basic}}                & 0.34      & 0.29         \\
\multicolumn{1}{c|}{\textit{Contextual}}           &    -        & 0.27         \\
\multicolumn{3}{c}{\textbf{GPT-3.5}}                                  \\
\multicolumn{1}{c|}{\textit{Prompt type}} & \textit{Contextual} & \textit{Personalized} \\ \hline
\multicolumn{1}{c|}{\textit{Basic}}                & 0.17       & 0.16         \\
\multicolumn{1}{c|}{\textit{Contextual}}           &   -         & 0.15        
\end{tabular}
\end{minipage}\hfill%
\begin{minipage}[t]{0.48\linewidth}\centering
\caption{Comparing the diversity of synthetic generated data using different input prompts by applying chrF score}
\label{Tabella_chrf}
\label{tab:The parameters 2 }
\begin{tabular}{ccc}
\multicolumn{3}{c}{\textbf{Phi-mini}}                                 \\
\multicolumn{1}{c|}{\textit{Prompt type}} & \textit{Contextual} & \textit{Personalized} \\ \hline
\multicolumn{1}{c|}{\textit{Basic}}                & 58.57       & 54.01         \\
\multicolumn{1}{c|}{\textit{Contextual}}           &    -        & 52.71       \\
\multicolumn{3}{c}{\textbf{Phi-medium}}                               \\
\multicolumn{1}{c|}{\textit{Prompt type}} & \textit{Contextual} & \textit{Personalized} \\ \hline
\multicolumn{1}{c|}{\textit{Basic}}                & 61.50       & 56.26         \\
\multicolumn{1}{c|}{\textit{Contextual}}           &    -        & 53.52       \\
\multicolumn{3}{c}{\textbf{GPT-3.5}}                                  \\
\multicolumn{1}{c|}{\textit{Prompt type}} & \textit{Contextual} & \textit{Personalized} \\ \hline
\multicolumn{1}{c|}{\textit{Basic}}                & 49.07       & 48.73         \\
\multicolumn{1}{c|}{\textit{Contextual}}           &    -        & 47.44        
\end{tabular}
\end{minipage}
\end{table*}
\subsection{Data Diversity}
We use BLEU \cite{Papineni02bleu} and chrF \cite{popovic-2015-chrf} to evaluate how much the answers generated with the three different prompts are similar with respect to $n$-grams overlapping. BLEU and chrF are two metrics commonly used for the Machine Translation task, the score of which is based on the comparison of $n$-gram overlap in tokens from the prediction and reference text. BLEU’s output is a number between 0 and 1, indicating how similar the candidate text is to the reference texts, with values closer to 1 representing similar texts. The chrF score can be any value between 0 and 100, inclusive. The results are in Tables \ref{Tabella_bleu} and \ref{Tabella_chrf}. For BLEU scores, the Phi-mini and Phi-medium models displayed the greatest diversity in generated texts when comparing \textit{Contextual} and \textit{Personalized} prompts outputs. 
GPT-3.5 shows lower BLEU scores overall: 
the BLEU score between answers generated with the \textit{Contextual} and the \textit{Personalized} prompts is 0.15, which is the lowest score obtained. The chrF scores supported these findings, with lower scores reflecting higher diversity. Both Phi-mini and medium exhibit the most variability with \textit{Contextual}-\textit{Personalized} prompts, while \textit{Personalized}-\textit{Basic} and \textit{Basic}-\textit{Contextual} prompts led to slightly more similar outputs. 
Even chrF scores show that GPT-3.5 produces the more diverse outputs compared to the others LLMS: 
The chrF score between the texts generated through the \textit{Contextual} and the \textit{Personalized} prompts is 47.44, which is the lower score obtained. Overall, these results highlight that all models produced slightly different answers depending on the prompt type, particularly between \textit{Contextual} and \textit{Personalized} prompts outputs, with GPT-3.5 exhibiting the greatest diversity in response generation. Finally, we analyse the lexical overlap between the Body of each question and both human-written and 
LLMs generated answers. The average percentages of query words that occur in the human-written answers is $23.4 \%$.
The average percentages of query words that occur in Phi-3-mini-generated responses are $34.0\%$, $35.4\%$ and $35.5\%$ for the \textit{Basic}, \textit{Contextual} and \textit{Personalized} prompt, respectively. The average percentages of query words in Phi-3-medium-generated answers are $35.4\%$, $35.5\%$ and $34.0\%$ for the \textit{Basic}, \textit{Contextual} and \textit{Personalized} prompt, respectively. The average percentages of query words that occur in ChatGPT-generated responses are $33.8\%$, $34.5\%$ and $33.2\%$ for the \textit{Basic}, \textit{Contextual} and \textit{Personalized} prompt, respectively. We suspect that this higher lexical overlap compared to the human answers happens because LLMs often repeat the question or query in the response \cite{askari_gpt}. It is worth noting that lexical overlap is not the best indicator of answer quality for fine-tuning effective encoders: since we evaluate our IR models on a human-written test set, there may be cases where human-written responses with low lexical overlap are relevant and informative, especially in question-answering tasks.

\subsection{Factual Information in generated answers}
While LLMs have shown human-like capabilities in generating text, they are prone to hallucination, producing plausible yet incorrect statements \cite{tonmoy2024comprehensive}. 
This, combined with potentially outdated information, undermines the reliability of synthetic answers. 
Relying on such information increases the risk of spreading misinformation, particularly in communities where accurate information is critical, such as health, legal or finance.
To assess the impact of hallucinations, we randomly select a total of 
100 questions equally split among all communities, and manually check the synthetic answers.\footnote{The subset of manually checked answers is available on the GitHub page (see Footnote 1).} We found that Phi-mini, Phi-medium, and GPT-3.5 generated incorrect answers in 41\%, 35\%, and 36\% of cases, respectively. 
In Table \ref{tab:confronto_history} we illustrate an example of hallucination. We ask the title of an Italian movie where a cartoonist is transformed into an animated character. Even if both Phi-mini and medium suggest the title of an Italian movie, none of the models and prompts generate a correct answer. The answer generated by GPT-3.5 with the \textit{Basic} and \textit{Personalized} prompts talks about "Mokeybone": even if "Monkeybone" is not an Italian movie but an American one, there are both live-action and animated characters, making it a plausible answer to the user's question. It is worth noting that GPT-3.5 with the \textit{Contextual} prompt generates produced an invented title, "Il lupo perde il pelo" (translated as "The Wolf Loses His Pur"), which does not correspond to any known movie, but it is a popular Italian saying. This observation highlights the model's capacity for generating both plausible but incorrect answers and fictional content. 
Despite these challenges, as shown in Section \ref{Results}, models trained on synthetic data can achieve comparable or even better retrieval performance compared to the models trained on human-written answers.
 




\section{Conclusion and Future Works}

In this study, we explore the generation of synthetic data for Personalized Information Retrieval (PIR) tasks using Large Language Models (LLMs). By leveraging the SE-PQA dataset, we generate synthetic answers and propose a new dataset, Sy-SE-PQA, with varying levels of personalization and contextual information, employing models like GPT-3.5 Turbo, Phi-3-mini, and Phi-3-medium. Our experiments show that neural re-rankers, in particular, DistillBERT, fine-tuned on synthetic data, 
outperformed the traditional BM25 method. This indicates that synthetic datasets can effectively train neural retrieval models, enhancing their ability to provide relevant and precise answers. Our findings highlight the potential of synthetic data in improving PIR tasks, especially when personalized and context-aware responses are crucial. The notable improvements in metrics over the baselines underscore the value of incorporating personalized information into the answer generation process. Furthermore, the exploratory analysis revealed challenges related to hallucinations in generated answers, emphasizing the need for continued refinement of LLMs to ensure accuracy and reliability. We leave as future work the development of other prompt techniques that could exploit a wider range of user-related and contextual features available in the SE-PQA dataset, which are not used in our current study. Additionally, incorporating Retrieval Augmented Generation (RAG) methods could be explored to reduce the occurrence of incorrect responses. 
By doing so, we hope to enhance the accuracy and relevance of generated answers, further improving the robustness of retrieval models.

\section{Limitations}
Despite the promising results, our study has several limitations that need to be discussed:


\paragraph{Bias and Fairness}
An ethical concern associated with the content generated with LLMs is bias and fairness. 
The LLMs are trained on large datasets that may contain biases reflecting societal prejudices. These biases can be, inadvertently, perpetuated by utilizing data generated by these LLMs, which could lead to unfair or discriminatory outcomes.


\paragraph{Computational and Budget Constraints} 
The generation of synthetic data using LLMs is resource intensive. Due to the computational and budget constraints, we limited the number of questions we could generate the answers to, affecting potentially the diversity of the generated dataset.

\paragraph{Evaluation Metrics} 
In this work we relied on the classical IR metrics such as P@1, NDCG@3, NDCG@10 and MAP@100, which certainly are useful to evaluate the retrieval effectiveness, but they may not capture completely the quality of the generated answers, for example, readability, user satisfaction and coherence.


While the synthetic data holds great potential, addressing these limitations is very important for advancing the field, while ensuring the robustness and applicability of the models in real-world scenarios.

\section*{Acknowledgment}
We acknowledge the CINECA award under the ISCRA initiative, for the availability of high-performance computing resources and support. This work was partially supported by the European Union – Next Generation EU within the project NRPP M4C2, Investment 1.,3 DD. 341 - 15 march 2022 – FAIR – Future Artificial Intelligence Research – Spoke 4 - PE00000013 - D53C22002380006.


\bibliographystyle{IEEEtran}
\bibliography{mybibfile}

\end{document}